\documentclass[a4paper,11pt]{article}
\usepackage{pos}

\title{Isovector Axial Vector Form Factors of the Nucleon from Lattice QCD with $N_{f}=2+1$ $\mathcal O(a)$-improved Wilson Fermions}
\ShortTitle{Isovector Axial Vector Nucleon Form Factors with $N_{f}=2+1$ Wilson Clover Fermions}

\author[a,b]{Dalibor Djukanovic}
\author[c]{Georg von Hippel}
\author[c]{Jonna Koponen}
\author[a,b,c]{Harvey B. Meyer}
\author[c]{Konstantin Ottnad}
\author*[c]{Tobias Schulz}
\author[a,b,c]{Hartmut Wittig}

\affiliation[a]{Helmholtz Institute Mainz, \\ 
Staudingerweg 18, D-55128 Mainz, Germany}

\affiliation[b]{GSI Helmholtzzentrum für Schwerionenforschung, \\
Planckstraße 1, D-64291 Darmstadt, Germany}

\affiliation[c]{PRISMA$^+$ Cluster of Excellence and Institute for Nuclear Physics, \\
Johannes Gutenberg University of Mainz, Johann-Joachim-Becher-Weg 45, D-55128 Mainz, Germany}

\emailAdd{d.djukanovic@him.uni-mainz.de}
\emailAdd{hippel@uni-mainz.de}
\emailAdd{jkoponen@uni-mainz.de}
\emailAdd{meyerh@uni-mainz.de}
\emailAdd{kottnad@uni-mainz.de}
\emailAdd{schulzt@uni-mainz.de}
\emailAdd{wittigh@uni-mainz.de}

\abstract{
We present the analysis of isovector axial vector nucleon form factors on a set of $N_f=2+1$ CLS ensembles with $\mathcal O(a)$-improved Wilson fermions and Lüscher-Weisz gauge action. The set of ensembles covers a pion mass range 
of $130-353\,$MeV with lattice spacings between $0.05\,$fm and $0.09\,$fm. In particular, the 
set includes a $L/a=96$ ensemble at the physical pion mass. For the purpose of the form factor extraction, we employ both the summed operator insertion method (summation method) and explicit two-state fits in order to account for excited-state contributions to the nucleon correlation functions. To describe the $Q^{2}$-behavior of the form factors, we perform $z$-expansion fits. Finally, we present HBChPT-inspired chiral and continuum extrapolations of the axial charge and radius.

}
\FullConference{%
 The 38th International Symposium on Lattice Field Theory, LATTICE2021
  26th-30th July, 2021
  Zoom/Gather@Massachusetts Institute of Technology
}
\begin{document}
\maketitle
\section{Introduction}
The axial structure of the nucleon is of particular interest from both the experimental and the theoretical point of view as, compared with the electromagnetic nucleon structure, it is much less well known phenomenologically. 
There are mainly three different experimental approaches to study the axial form factors which parametrize the non-perturbative interaction vertex of 
the $W^{\pm}$ bosons and the nucleon. The relevant interaction processes comprise (in)elastic neutrino nucleon scattering $\nu_\mu+n\to \mu^-+p$, $\bar\nu_\mu+p\to \mu^++n$, 
(charged) pion electroproduction $\gamma^{*}+N\to N'+\pi$ and 
ordinary (or radiative) muon capture processes $\mu^-+p\to n+\nu_\mu\,(+\gamma)$. It is widely believed that the analysis of interactions between weakly interacting particles and ordinary matter represents a major perspective for dark matter searches. Various experimental collaborations such as DUNE, Hyper-Kamiokande, MiniBooNE and MINOS for instance are studying neutrino reactions with a focus on oscillation processes, mass hierarchies, CP violation and scattering cross sections. 
Of particular interest in this respect are the axial charge $g_{A}$ and 
the (mean-square) axial radius $\langle r_{A}^{2}\rangle$ of the nucleon. While the former is precisely ($0.1\%$ accuracy) known from cold neutron $\beta$-decay 
experiments yielding $g_{A}=1.2756(13)$ \cite{particleDataGroupLatest} the latter exhibits a discrepancy between the world average results from neutrino scattering $\langle r_{A}\rangle=0.666(14)\,\mathrm{fm}$ \cite{axialStructureNucleon} and pion electroproduction $\langle r_{A}\rangle=0.639(10)\,\mathrm{fm}$ \cite{axialStructureNucleon}. 
The more recent result quoted by the MiniBooNE collaboration is given 
by $\langle r_A\rangle=0.506(64)$ \cite{miniboone2010}. The situation concerning the axial radius determination is much more complex in comparison to the axial charge measurement due to the experimental challenges encountered when measuring scattering cross-sections at moderate energies. In addition, the 
typical $Q^{2}$-parametrization of the axial form factor using a dipole shape introduces a specific model dependence which is thought to underestimate the real uncertainty of the form 
factor \cite{Bodek2008,refmeyer2016,refhill2018}. 

We present our results of a state-of-the-art analysis of the isovector axial vector nucleon form factor based on lattice data produced with $N_{f}=2+1$ $\mathcal{O}(a)$-improved Wilson fermions on a set of CLS gauge ensembles. We employ both the summation method and explicit two-state fits to extract the ground-state contribution. The $Q^{2}$-dependence of the form factor is parametrized using the $z$-expansion and the subsequent chiral extrapolation of the extracted axial radius is performed using the Heavy Baryon Chiral Perturbation Theory (HBChPT) formalism. Our final results are obtained from a model average procedure based on the Akaike information criterion including various different model variations. 
\section{Methodology}
We start our discussion by considering the decomposition of the nucleon matrix element of the isovector axial vector current, which is based on Lorentz invariance and isospin symmetry:
\begin{align}
\label{eq:axialmatrixelement}
\langle N(\vec p',s')|A_\mu^{a}(0)|N(\vec p,s)\rangle
=
\bar{u}(\vec p',s')\left[
\gamma_\mu\gamma_5 G_A(Q^2)-i\gamma_5\frac{Q_\mu}{2m_N}G_P(Q^2)
\right]\frac{\tau^a}{2}u(\vec p,s),
\end{align}
with $u(\vec p,s)$ and $Q_{\mu}=(iq_0,\vec q)$ denoting the nucleon Dirac spinor in momentum space, and the Euclidean four-momentum $Q^{2}=-q^{2}$. In this case, $G_{A}(Q^2)$ and $G_{P}(Q^2)$ denote the axial and the induced pseudoscalar form factors, respectively. We construct the following ratio of nucleon 3-point and 2-point correlation functions \cite{refratio1,refratio2}
\begin{align}
\label{eq:RatioCorrelationFunctions}
R_{A_\mu^a}(\Gamma;\vec p',\vec p,t_s,t)
&=
\frac{C_{3,A_\mu^a}^N(\Gamma;\vec p',\vec p,t_s,t)}{C_2^N(\Gamma;\vec p',t_s)}
\sqrt{\frac{C_2^N(\Gamma;\vec p,t_s-t)C_2^N(\Gamma;\vec p',t)C_2^N(\Gamma;\vec p',t_s)}
{C_2^N(\Gamma;\vec p',t_s-t)C_2^N(\Gamma;\vec p,t)C_2^N(\Gamma;\vec p,t_s)}},
\end{align}
with
\begin{align}
\label{eq:nucleon_2pt}
C_2^N(\Gamma;\vec p,t)
&=
a^3\sum_{\vec x}e^{-i\vec p\cdot\vec x}\,
\Gamma_{\alpha\beta}\langle\mathcal O_N(\vec x,t)_\beta
\mathcal{\bar O}_N(\vec 0,0)_\alpha
\rangle,
\\
\label{eq:nucleon_3pt}
C_{3,A_\mu^a}^N(\Gamma;\vec p',\vec p,t_s,t)
&=
a^6\sum_{\vec x,\vec y}
e^{-i\vec p'\cdot(\vec x-\vec y)-i\vec p\cdot\vec y}
\Gamma_{\alpha\beta}
\langle 
\mathcal O_N(\vec x,t_s)_\beta
A_\mu^a(\vec y,t)
\mathcal{\bar O}_N(\vec 0,0)_\alpha
\rangle
\end{align}
in order to cancel the dependence on the time separations and the overlap factors. The spin and parity projection matrix is chosen to be ${\Gamma=(1+\gamma_0)(1+i\gamma_5\gamma_3)/2}$. In the notation given, $t_{s}$ and $t$ denote the source-sink separation and the operator insertion time, respectively. Furthermore, the kinematics are chosen such that the nucleon at the sink is at rest, $\vec p'=0$, yielding $\vec q=\vec p'-\vec p=-\vec q$. 

In the so-called asymptotic limit $t\gg a,(t_s-t)\gg a$, the ratio given 
in eq. (\ref{eq:RatioCorrelationFunctions}) is expected to be dominated by the corresponding ground state contribution $R(\Gamma;\vec q)$ which is derived by inserting the spectral representations of the correlation functions given in eqs. (\ref{eq:nucleon_2pt}) and (\ref{eq:nucleon_3pt}) 
into eq. (\ref{eq:RatioCorrelationFunctions}) yielding
\begin{align}
\label{eq:ratioComponentsAxialTemporal}
R_{A_0}(\Gamma;Q^2,t_s,t)
&=
\frac{q_3}{\sqrt{2E_N(\vec q)(E_N(\vec q)+m_N)}}
\left(
G_A(Q^2)+\frac{m_N-E_N(\vec q)}{2m_N}G_P(Q^2)
\right)+\dots,
\\
\label{eq:ratioComponentsAxialSpatial}
R_{A_j}(\Gamma;Q^2,t_s,t)
&=
\frac{i}{\sqrt{2E_N(\vec q)(E_N(\vec q)+m_N)}}
\left(
(E_N(\vec q)+m_N)G_A(Q^2)\delta_{j3}-\frac{q_jq_3}{2m_N}G_P(Q^2)
\right)+\dots.
\end{align}
The ellipses represent higher-order (exponentially suppressed) excited states. For every fixed value of the momentum transfer $Q^{2}$ we solve 
the (overdetermined) linear system of equations $\vec R=M\vec G$ for the spatial ratio 
components $R_{A_j}$ given in eq. (\ref{eq:ratioComponentsAxialSpatial}) for each $t_{s}$ and $t$ by a least squares procedure minimizing the following $\chi^{2}$ functional
\begin{align}
\label{eq:axial_linear_system}
\chi^2=(\vec R-M\vec G)^T\mathrm{Cov}^{-1}(\vec R-M\vec G),\quad
\vec R=
\begin{pmatrix}
R_1 \\ R_2 \\ \dots \\R_N
\end{pmatrix},\quad
M=\begin{pmatrix}
M_1^A & M_1^P \\ 
\vdots & \vdots \\
M_N^A & M_N^P 
\end{pmatrix},\quad
\vec G=
\begin{pmatrix}
G_A \\ G_P
\end{pmatrix},
\end{align}
which allows the extraction of the effective axial form 
factor $G_{A}^{\mathrm{eff}}(Q^2,t_s,t)$. In this case, $N$ denotes the number of ratio components after averaging over equivalent lattice momenta at a given $Q^{2}$. We do not include the temporal ratio component $R_{A_0}$ in our analysis. In order to extract the ground-state contribution to $G_{A}(Q^2)$ we employ two different methods:
\\ \\
1. \textit{Summation Method:} The summed effective axial form factor $S(Q^2,t_s)$ is modeled with a linear function
\begin{align}
S(Q^2,t_s)=\sum_{t=a}^{t_s-a}G_{A}^{\mathrm{eff}}(Q^2,t_s,t)\to
\mathrm{const.}(Q^2)+G_A(Q^2)\frac{t_s}{a}+\mathcal{O}(t_se^{-\Delta(Q^2)t_s})+\mathcal{O}(t_se^{-\Delta(0)t_s})
\end{align}
where the extracted slope represents an estimate for the ground-state contribution. The exponential suppression of higher-order states 
depends on $t_{s}$ as well as on the (momentum dependent) energy gap between the ground and the first excited state, $\Delta(Q^2):=E_{1}(Q^2)-E_{0}(Q^2)$.
\\ \\
2. \textit{Two-State Fit:} We employ the following (two-state) model
\begin{equation}
\begin{aligned}
G_A^{\text{eff}}(Q^2,t_s,t)
% R_{\mathcal J}(\vec p',\vec p,t_s,t)
=\ &
% k_{00}(\vec p',\vec p)
G_A(Q^2)
\Big[
1+k_{01}(Q^2)e^{-\Delta(Q^2)t}+k_{10}(Q^2)
e^{-\Delta(0)(t_s-t)}+k_{11}(Q^2)e^{-\Delta(0)(t_s-t)-\Delta(Q^2)t}+
\\
&
\begin{aligned}
&
+\frac{1}{2}\tilde a_1(Q^2)
\left(e^{-\Delta(Q^2)(t_s-t)}-e^{-\Delta(Q^2)t_s}\right)+
\frac{1}{2}\tilde a_1(0)\left(e^{-\Delta(0)t}-e^{-\Delta(0)t_s}\right)
\Big]+\dots
\end{aligned}
\end{aligned}
\end{equation}
which explicitly takes the contribution of the first excited state into account. At every fixed value of $Q^{2}$ we fit all data points for $G_A^{\text{eff}}(Q^2,t_s,t)$ simultaneously including all values $t\geq t_{\mathrm{start}}$ included in the interval centered around $t_{s}/2$. Furthermore, we add a Gaussian prior for the energy gaps $\Delta(0),\Delta(Q^2)$ and the overlap 
factors $\tilde a_{1}(0),\tilde a_{1}(Q^2)$ where the corresponding prior 
means $p$ and prior widths $p_w$ are set based on prior knowledge obtained from results of separate two-state fits to the nucleon 2-point correlation function. We follow a specific procedure to allow for a larger prior width if the statistical error of the fit result as a function of the prior width is bounded by a certain factor.

We have performed explicit $\mathcal{O}(a)$-improvement and non-perturbative renormalization of the axial current,
\begin{align}
(A^a_\mu)^I_R(x)&=
Z_A(\tilde g_0^2)(1+b_A(g_0^2)am_l+\bar{b}_A(g_0^2)(2am_l+am_s))
\left(
A^a_\mu(x)+ac_A(g_0^2)\tilde \partial_\mu P^a(x)
\right)
,
\end{align}
% \begin{align}
% (A^a_\mu)_I(x)=A^a_\mu(x)+ac_A(g_0^2)\tilde \partial_\mu P^a(x),
% \end{align}
with $\tilde{\partial}_{\mu}$, $P^a(x)$ and $c_{A}$ denoting the central discrete lattice derivative, the pseudoscalar density and the axial improvement 
coefficient $c_{A}$ (taken from \cite{improvementmainz}), respectively. 
% As for the non-perturbative renormalization 
% \begin{align}
% (A^a_\mu)_R(x)&=
% Z_A(\tilde g_0^2)(1+b_A(g_0^2)am_l+\bar{b}_A(g_0^2)(2am_l+am_s)),
% \end{align}
We do not include the contribution proportional to $\bar b_{A}$ (which is of order $\mathcal O(g_0^4)$ in lattice perturbation theory). The values 
for $b_{A}$ and $Z_{A}$ are taken from \cite{renormalizationb} and \cite{renormalizationz}, respectively.
\section{Lattice Setup}
Our analysis comprises twelve CLS gauge ensembles which have been generated 
with $N_{f}=2+1$ dynamical $\mathcal O(a)$-improved Wilson fermions and the tree-level improved Lüscher-Weisz gauge action \cite{scaleSetting2015,scaleSetting2017}. The quark masses have been tuned to follow the chiral trajectory of ${\mathrm{Tr}(M)=2m_{l}+m_{s}=\mathrm{const.}}$. For the majority of ensembles the fields obey open boundary conditions in the time direction in order to prevent topological freezing. In table \ref{tab:cls_table} we show an overview of lattice geometries and hadron masses for the ensembles included in the analysis.
\begin{table}[htp]
\begin{center}
\begin{tabular}{cccccccc}
\hline
Ensemble & $\beta$ & $L/a$ & $T/a$ & $M_\pi$ [MeV] & $M_\pi L$ & $m_N$ [MeV] & $t_{s}$ [fm] \\
\hline
H105 & 3.40 & 32 & 96 & $278$ & $3.89$ & $1020$ & $1.0, 1.2, 1.4$ \\
% \hline
C101 & 3.40 & 48 & 96 & $223$ & $4.68$ & $984$ & $1.0, 1.2, 1.4$ \\
\hline
S400 & 3.46 & 32 & 128 & $350$ & $4.33$ & $1123$ & $1.1, 1.2, 1.4, 1.5, 1.7$ \\
% \hline
N451$^{*}$ & 3.46 & 48 & 128 & $286$ & $5.31$ & $1049$ & $1.2, 1.4, 1.5$ \\
% \hline
D450$^{*}$ & 3.46 & 64 & 128 & $216$ & $5.36$ & $966$ & $1.1, 1.2, 1.4, 1.5$ \\
\hline
S201 & 3.55 & 32 & 128 & $293$ & $3.05$ & $1098$ & $1.0, 1.2, 1.3, 1.4$ \\
% \hline
N203 & 3.55 & 48 & 128 & $347$ & $5.42$ & $1105$ & $1.0, 1.2, 1.3, 1.4, 1.5$ \\
% \hline
N200 & 3.55 & 48 & 128 & $283$ & $4.42$ & $1053$ & $1.0, 1.2, 1.3, 1.4$ \\
% \hline
D200 & 3.55 & 64 & 128 & $203$ & $4.23$ & $960$ & $1.0, 1.2, 1.3, 1.4$ \\
% \hline
E250$^{*}$ & 3.55 & 96 & 192 & $130$ & $4.06$ & $938$ & $1.0, 1.2, 1.3, 1.4$ \\
\hline
N302 & 3.70 & 48 & 128 & $353$ & $4.28$ & $1117$ & $1.0, 1.1, 1.2, 1.3, 1.4$ \\
% \hline
J303 & 3.70 & 64 & 192 & $262$ & $4.23$ & $1052$ & $1.0, 1.1, 1.2, 1.3$ \\
\end{tabular}
\end{center}
\caption{Table showing the list of CLS gauge ensembles which are included in the analysis with information about the lattice geometries, pion and nucleon masses (at present level of statistics) as well as the list of source-sink separations used for the nucleon 3-point correlation function 
(see also \cite{refMainz2021}). The ensembles with periodic boundary conditions in the time direction are marked with an asterisk.}
\label{tab:cls_table}
\end{table}
The nucleon interpolating operator is constructed using Gaussian smeared quark fields where the smearing parameters have been tuned to realize a smearing radius of $r_{G}\approx 0.5\,\mathrm{fm}$. In addition, APE-smearing has been employed for the link variables.
The scale setting has been performed with the Wilson gradient flow scale which is explicitly given 
by $\sqrt{8t_0^{\mathrm{phys}}}=0.415(4)(2)\,\mathrm{fm}$ \cite{scaleSetting2017}.
\section{Modelling of the $\mathbf{Q^{2}}$-dependence and Extrapolations}
\begin{figure}[htp]
\centering
\includegraphics[width=0.45\linewidth]{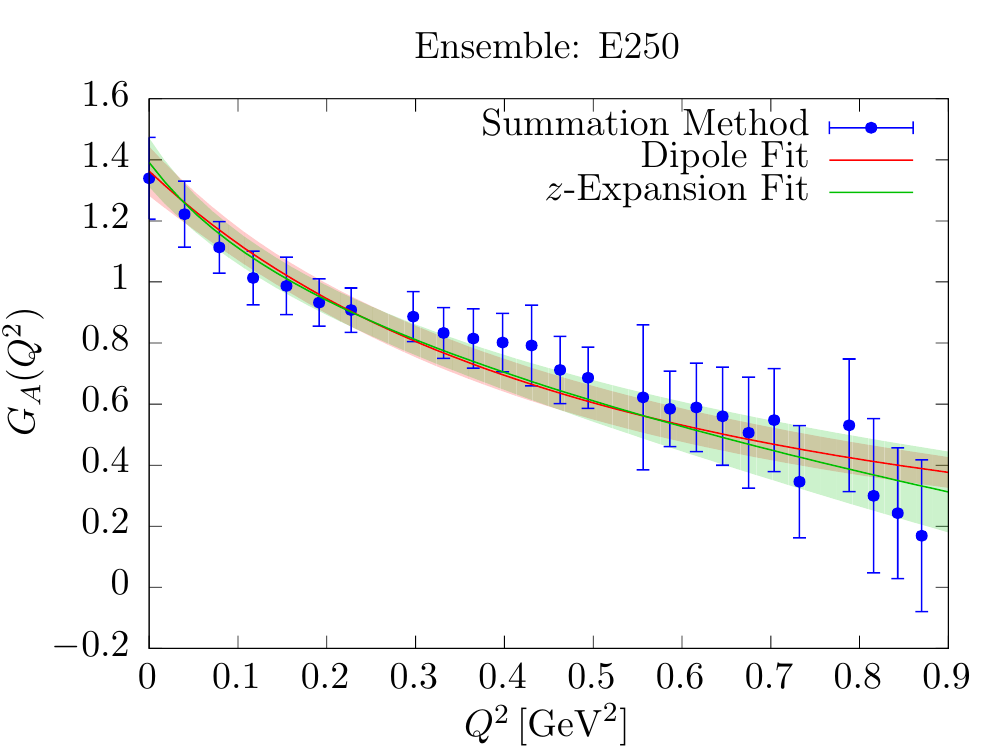}
\includegraphics[width=0.45\linewidth]{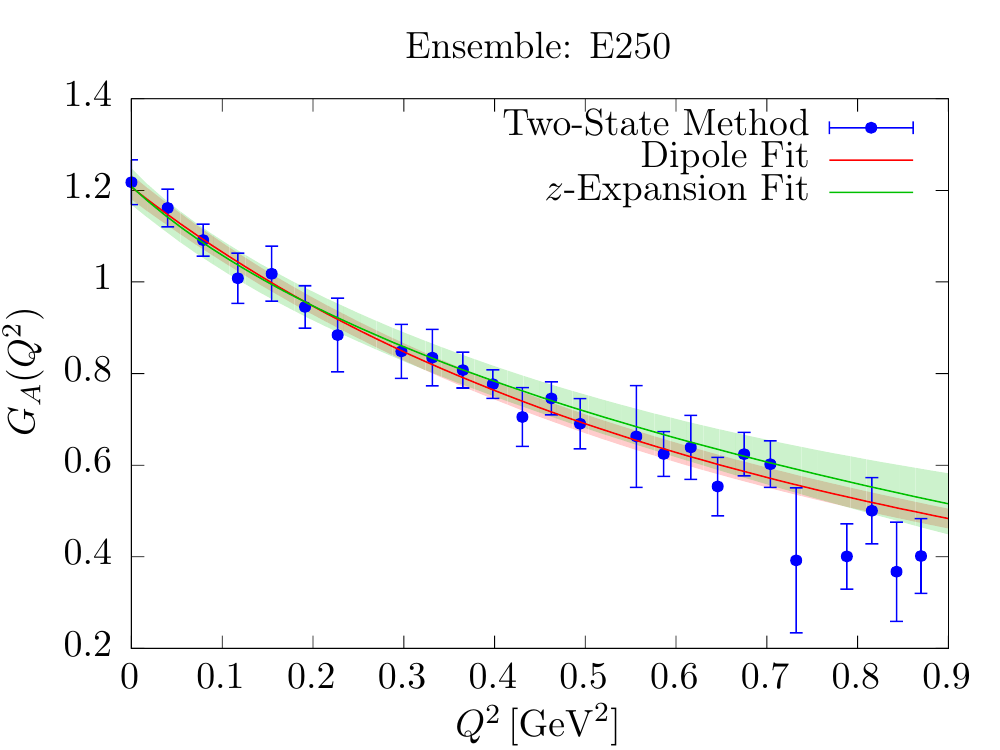}
\caption{Results of the $z$-expansion fits (green curve and error band) on the E250 ensemble with ${Q^{2}\leq 0.9\,\mathrm{GeV}^{2}}$ for the summation method (left) and the two-state fits (right). Additionally, the corresponding dipole fits are shown for the sake of comparison.}
\label{fig:z_expansion_e250}
\end{figure}

We employ the model independent $z$-expansion 
parametrization \cite{zExpansionReference} of the axial form factor given by
\begin{align}
G_{A}(Q^2)=
\sum_{k=0}^N
a^{A}_k z^k(Q^2),\quad
z(Q^2)=
\frac{\sqrt{t_\mathrm{cut}+Q^2}-\sqrt{t_\mathrm{cut}}}
{\sqrt{t_\mathrm{cut}+Q^2}+\sqrt{t_\mathrm{cut}}},
\end{align}
where we have chosen to map the point $Q^2=0$ to $z=0$ ($t_{0}=0$). In addition, the value for the cut on the real axis is given by $t_{\mathrm{cut}}=9M_{\pi}^{2}$ for the isovector axial vector channel. For our choice of $t_{0}$ the real expansion 
parameters $\{a_k^A\}_{k=1,2,\dots}$ are related to the axial charge and radius in the following way
\begin{align}
g_A=G_A(0)=a^A_0,\quad 
\langle r_A^2\rangle=-\frac{6}{G_A(0)}\frac{\partial G_A(Q^2)}{\partial Q^2}
\Bigg|_{Q^2=0}=-\frac{3a_1^A}{2a_0^At_{\mathrm{cut}}}.
\end{align}
We restrict higher order terms in the expansion from reaching unphysically large values by adding Gaussian priors for all 
coefficients $\{a_k^A\}_{k\geq 2}$. Following this procedure we find that a total number of five expansion terms ($N=4$) is sufficient to obtain stable fit results.  
In order to check for the momentum transfer dependence, we employ several momentum cuts given by $Q^{2}_{\mathrm{cut}}\in\{0.6,0.7,0.8,0.9\}\,\mathrm{GeV}^{2}$. 
In fig. \ref{fig:z_expansion_e250} we show a subset of results of 
the $z$-expansion fits on the E250 ensemble ($M_{\pi}\approx 130\,\mathrm{MeV}$) with $Q^{2}\leq 0.9\,\mathrm{GeV}^{2}$ for both the summation method (left panel) with $\chi^{2}/\mathrm{d.o.f.}=0.94$, $p$-value $=0.54$, and the two-state fit (right panel) with $\chi^{2}/\mathrm{d.o.f.}=0.38$, $p$-value $\approx 1.00$.
We want to put our main focus on the axial radius extraction so we decide to perform the $z$-expansion fits and the subsequent extrapolation on the normalized effective form factor data for the summation and the two-state extractions corresponding to $G_{A}^{\mathrm{eff}}(Q^2,t_s,t)/G_{A}^{\mathrm{eff}}(0,t_s,t)$. In this case the $z$-expansion fits do not include the data points at $Q^{2}=0$ but constrain the zeroth order coefficient to $a_{0}^{A}=1$.  

\begin{figure}[htp]
\centering
\includegraphics[width=0.45\linewidth]{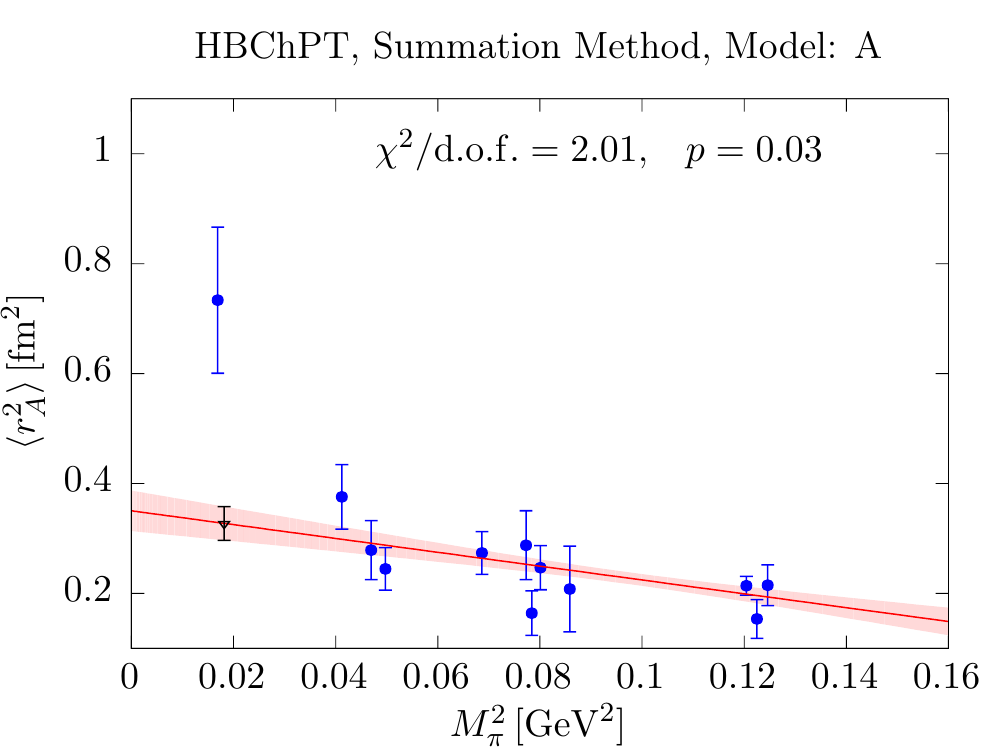}
\includegraphics[width=0.45\linewidth]{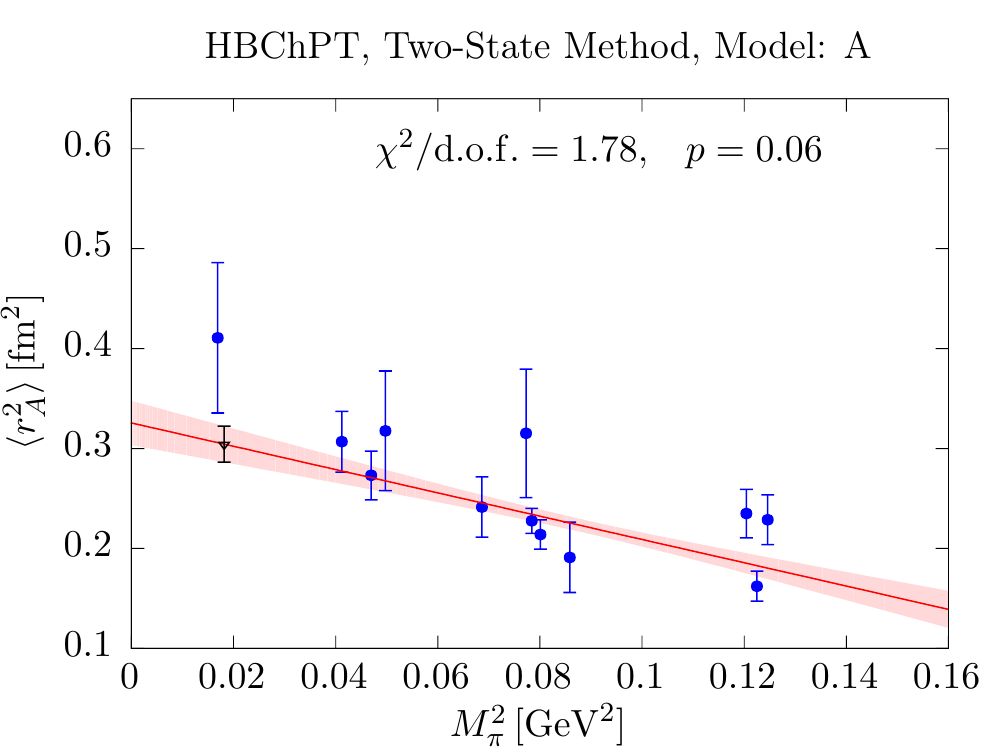}
\includegraphics[width=0.45\linewidth]{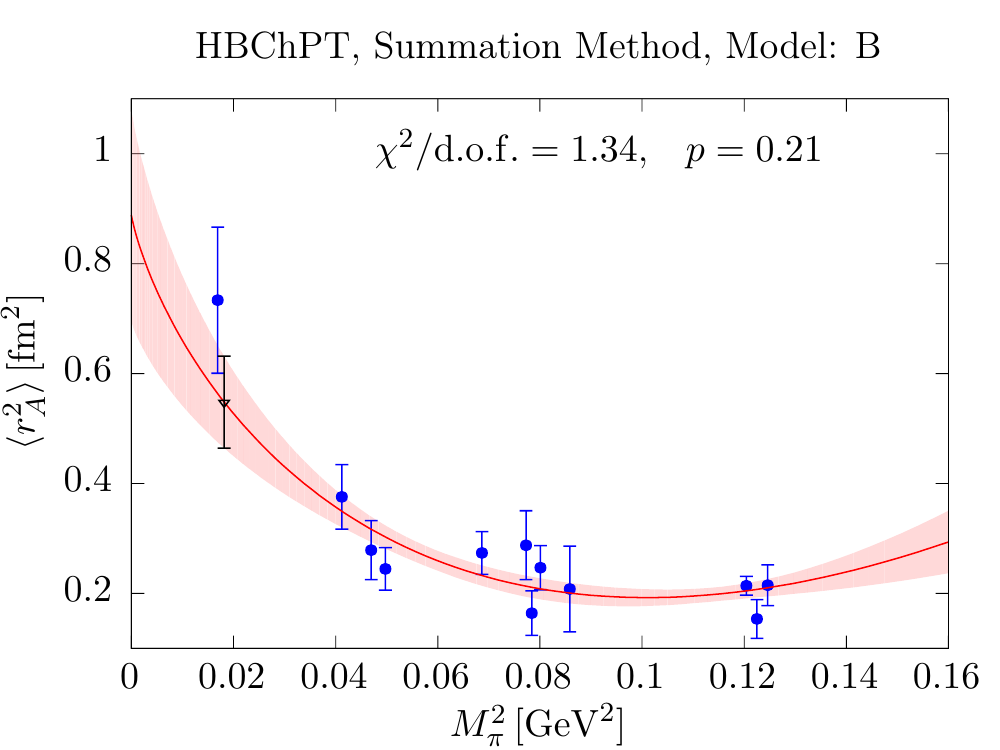}
\includegraphics[width=0.45\linewidth]{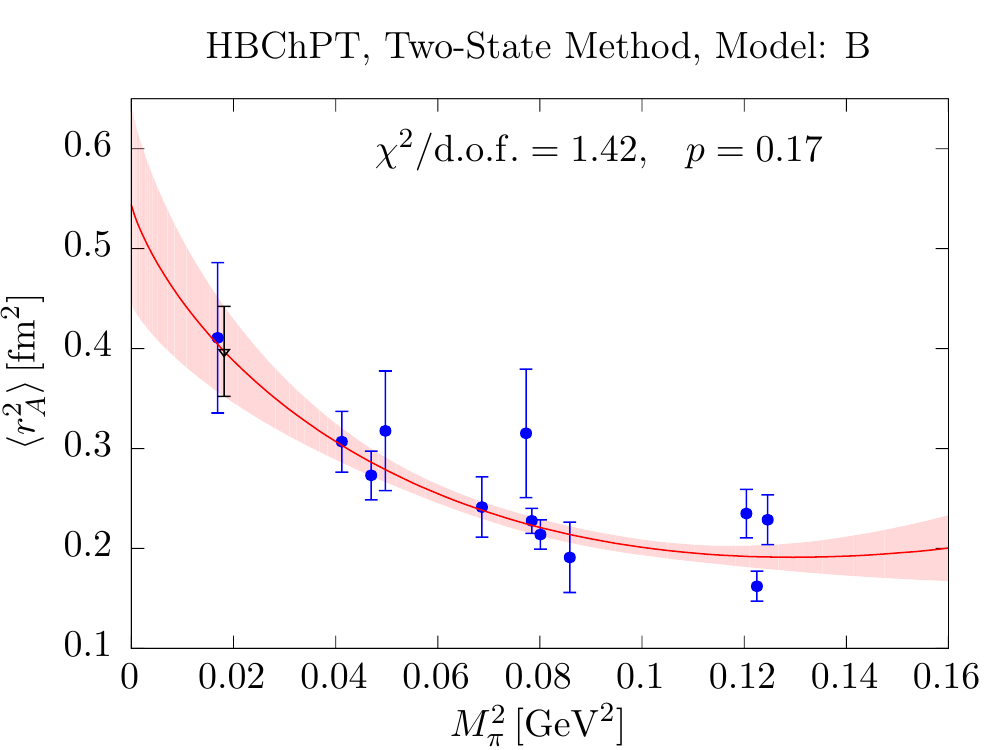}
\caption{Chiral extrapolation of the summation method (left column) and two-state fit (right column) results for both chiral models $A$ (top) and $B$ (bottom) with $Q^{2}_{\mathrm{cut}}=0.9\,\mathrm{GeV}^{2}$, $M_\pi^{\mathrm{cut}}=360\,\mathrm{MeV}$ and without including cutoff and finite volume terms.} 
% with $Q^{2}_{\mathrm{cut}}=0.9\,\mathrm{GeV}^{2}$, $M^\pi_{\mathrm{cut}}=360\,\mathrm{MeV}$.}
\label{fig:extrapolations_a_b}
\end{figure}

We perform a combined chiral, continuum and infinite-volume extrapolation of our results 
for $\langle r_{A}^2\rangle$ obtained from the $z$-expansion fits on every ensemble. With regard to the chiral extrapolation $M_{\pi}\to M_{\pi}^{\mathrm{phys}}$ we perform fits which are based on 
Heavy Baryon Chiral Perturbation Theory (HBChPT) \cite{hbchpt1} and use the physical pion mass estimate of $M_{\pi}^{\mathrm{phys}}=134.8(3)\,\mathrm{MeV}$ neglecting electromagnetic and isospin breaking effects \cite{flag2017}. For observable ${\cal O}$, we employ two different models which differ in the chiral order 
\begin{align}
\label{eq:extrapolation_axial_a}
&A:\ \mathcal O(M_\pi,a,L)=A_{\mathcal O}+B_{\mathcal O}M_\pi^2+D_{\mathcal O}a^2+E_{\mathcal O}M_\pi^2e^{-M_\pi L},
\\
\label{eq:extrapolation_axial_b}
&B:\ \mathcal O(M_\pi,a,L)=A_{\mathcal O}+B_{\mathcal O}M_\pi^2+C_{\mathcal O}M_\pi^2\log(M_\pi)+D_{\mathcal O}a^2+E_{\mathcal O}M_\pi^2e^{-M_\pi L}.
\end{align}
Furthermore, we employ two pion-mass cuts given by $M_{\pi}^{\mathrm{cut}}\in\{300,360\}\,\mathrm{MeV}$.
In fig. \ref{fig:extrapolations_a_b} we show the summation method (left column) and the two-state 
fit (right column) results for both models $A$ (top) and $B$ (bottom) without including cutoff and finite-volume terms $D_{\mathcal O}=0$, $E_{\mathcal O}=0$.
We find the two-state fit results to be more stable with respect to the inclusion of cutoff and finite volume terms. Since we find unexpectedly 
large ($\sim 40\%$) cutoff effects for the summation results we employ a Gaussian prior for $D_{\mathcal O}$ restricting the size of the coefficient such that cutoff effects are limited to at most $10\%$ for the coarsest lattice spacing. Additionally, the summation results for fit model $B$ show unreasonably large values for the coefficients indicating an artificial behavior of the fit. For this reason we add another prior for the coefficient $C_{\mathcal O}$ based on the size of the analytical result for the coefficient $C_{\mathcal O}$ in the case of the axial charge \cite{hbchpt1}. 
\section{Final Results and Outlook}
\begin{figure}[htp]
\centering
\includegraphics[width=0.49\linewidth]{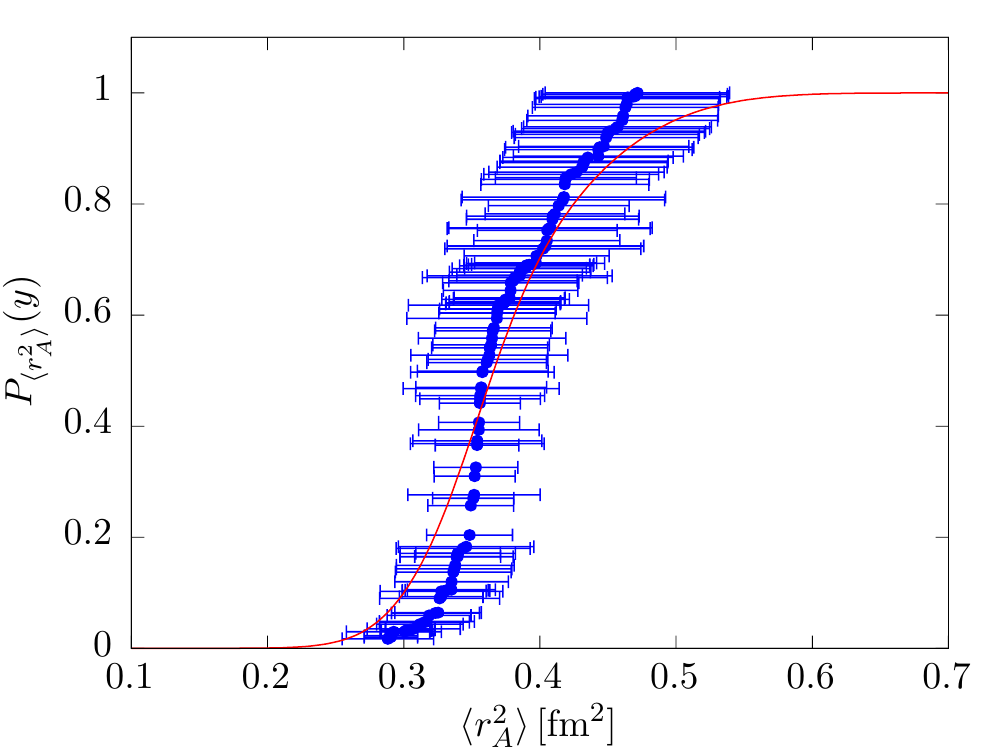}
\includegraphics[width=0.35\linewidth]{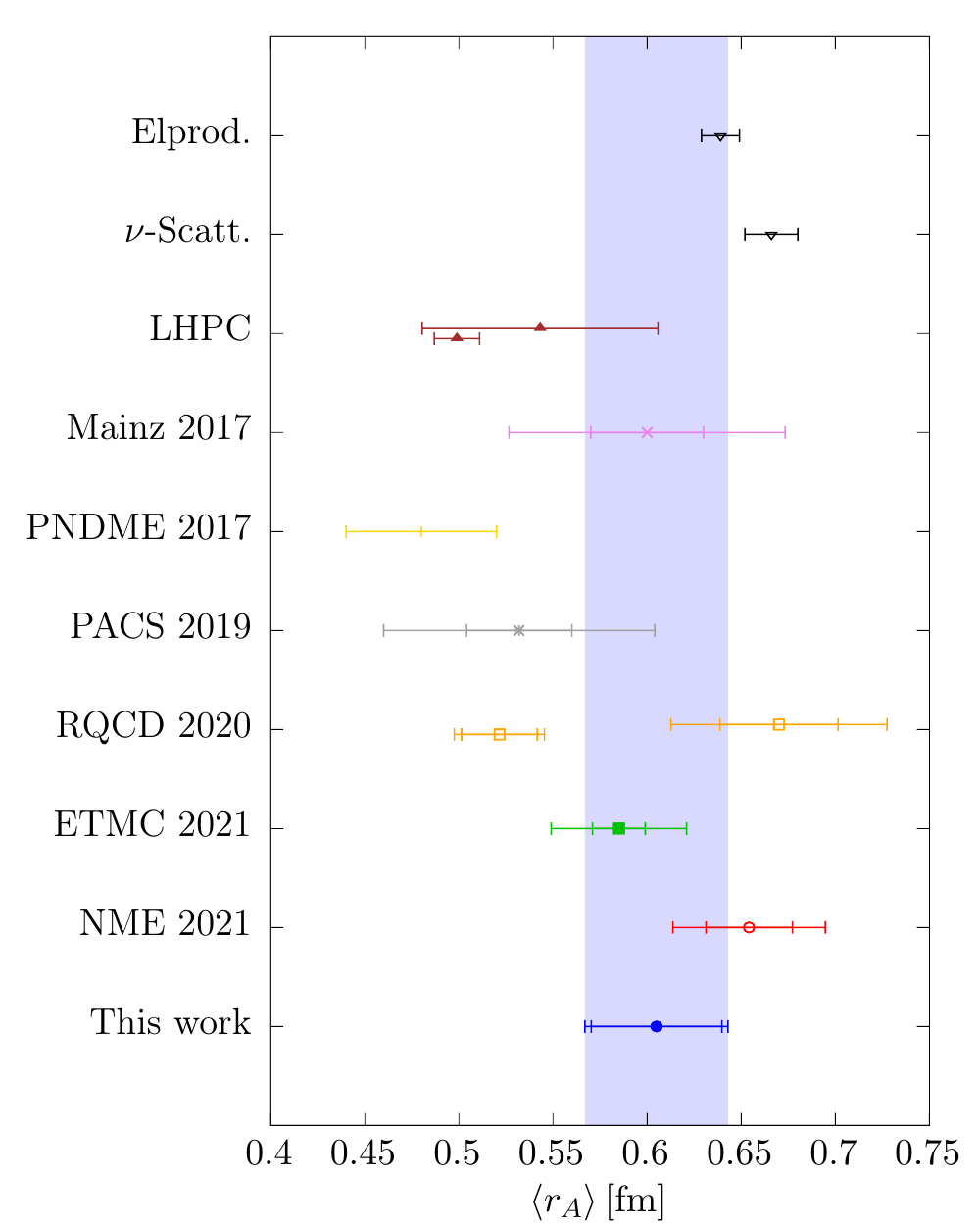}
\caption{Left: Combined weighted cumulative distribution function (red curve) of all model 
variations (blue data points). Right: Comparison of our final result with those obtained from  experiment and various lattice collaborations.}
\label{fig:final_results}
\end{figure}

We perform a model average procedure based on the Akaike information 
criterion \cite{akaike1974,jayNeil2021,borsanyi2021} associating an AIC-weight to every variation of the extrapolation defined 
by $\mathrm{AIC}:=\chi^{2}_{\mathrm{min}}+2n_{F}+2n_{C}$ with $n_{F}$ and $n_{C}$ denoting the number of fit parameters and the number of cut data points, respectively. The normalized AIC-weights are used to construct a combined cumulative distribution function (CDF) given by
\begin{align}
\label{eq:cdf_ra}
P_{\langle r_A^2\rangle}(y)=\sum_iw_i^{\text{AIC}}F_{\mathcal N}(y;x_i,\sigma_i),
\quad
w_i^{\mathrm{AIC}}=\frac{e^{-\mathrm{AIC}_i/2}}{\sum_je^{-\mathrm{AIC}_j/2}}
\end{align}
including all model variations (summation/two-state, model $A/B$, 
$M^{\mathrm{cut}}_{\pi}$, $Q^2_{\mathrm{cut}}$, \ldots) represented by normal-distributed random variables with single CDFs denoted by $F_{\mathcal N}$. On the left-hand side of fig. \ref{fig:final_results} we show the combined CDF defined in eq. (\ref{eq:cdf_ra}). We consider the median of the corresponding distribution as the mean of our final result. The statistical and systematic error are obtained from the spread of the $1\sigma$ percentiles and the scaling behavior of the statistical errors of the model variations. 
As our final result we quote
\begin{align}
\langle r_A^2\rangle=0.366(46)_{\mathrm{stat}}(42)_{\mathrm{sys}}\,\mathrm{fm}^2,
\end{align}
which exhibits a statistical (systematic) error of $13\%$ ($11\%$). On the right-hand side of fig. \ref{fig:final_results} we show the comparison of our final result with the experimental values and those obtained by various other lattice 
collaborations \cite{pndme2017,mainz2019,pacs2019,regensburg2020,pndme2021axial,etmcAxial2021}. Within 
the $1\sigma$ error interval we are in good agreement with the results of other collaborations  and with the experimental world average for the electroproduction.

With regard to further analyses we will increase the number of gauge configurations on the E250 ensemble. Additionally, we plan to add a further gauge ensemble at $M_{\pi}\approx 160\,\mathrm{MeV}$ which will improve the situation concerning the chiral extrapolation. We also plan 
to analyze $G_{P}(Q^2)$ with a special focus on excited-state contributions which we assume to play the major role with respect to the systematic uncertainty.
%
% \\ \\
%
% \textbf{Acknowledgements:}
\acknowledgments
This research is partly supported by the Deutsche Forschungsgemeinschaft (DFG,
German Research Foundation) through the CRC SFB 1044 (DFG grant HI 2048/1-2 (Project No. 399400745)), and through the Cluster of Excellence (PRISMA+ EXC 2118/1) funded by the DFG (Project ID 39083149). 
This work is supported by the European Research Council (ERC) under the European Union's Horizon 2020 research and innovation program through grant agreement No. 771971-SIMDAMA. Calculations for this project were partly performed on the HPC clusters ``Clover'' and ``HIMster2'' at the Helmholtz 
Institute Mainz, and ``Mogon 2'' at JGU Mainz. 
Additional computer time has been allocated through projects HMZ21, HMZ23 and HMZ36 on ``JUQUEEN'' and ``JUWELS'' at NIC, J\"ulich. 
The authors also gratefully acknowledge the Gauss Centre for Supercomputing e.V.
(www.gauss-centre.eu) for funding this project by providing computing time on
HAZEL HEN at H\"ochstleistungsrechenzentrum Stuttgart
(www.hlrs.de) under project GCS-HQCD. We are grateful to
our colleagues within the CLS initiative for sharing gauge field configurations.
%
%
% \begin{itemize}
% \item Perform a \textbf{model average} ($\{\text{Summation,Two-State},A,B,Q^2_{\text{cut}},\mathcal O(a^2),\dots\}$) based on 
% the Akaike information criterion. 
% \item Define (normalized) AIC weights for the $i$-th model estimate given by
% \begin{align*}
% \text{AIC}:=\chi^2_{\text{min}}+2n_F+2n_C,\quad
% w_i^{\text{AIC}}=\frac{e^{-\text{AIC}_i/2}}{\sum_k e^{-\text{AIC}_k/2}}
% \end{align*}
% $n_{F}:$ number of fit parameters, $n_{C}:$ number of cut data points
% \vfill
% \item \textbf{Cumulative distribution function (CDF)}
% \begin{align*}
% P_{\langle r_A^2\rangle}(y)=\sum_iw_i^{\text{AIC}}F_{\mathcal N}(y;x_i,\sigma_i),\quad \sigma^2_{\text{tot}}=\left((y_{84}-y_{16})/2\right)^2=
% \sigma^2_{\text{stat}}+\sigma^2_{\text{sys}},\quad \text{s.t.:}\, P_{\langle r_A^2\rangle}(y_{16})=0.16
% \end{align*}
% \end{itemize}
% 
%
%
% Comparison with \cite{pndme2017,mainz2019,pacs2019,regensburg2020,pndme2021axial,etmcAxial2021}
%
% \section{Summary and Conclusions}
%
% \\ \\
% \noindent
% \textbf{Acknowledgements:} This research is supported by the Deutsche Forschungsgemeinschaft (DFG, German Research Foundation) through the Collaborative Research Center SFB 1044 “The low-energy frontier of the Standard Model,” under DFG Grant No. HI 2048/1-2 (Project
% No. 399400745), and in the Cluster of Excellence “Precision Physics, Fundamental Interactions and the DFG within the German Excellence strategy (Project ID Structure of Matter” (PRISMA + EXC 2118/1) funded by 39083149. Calculations for
% this project were partly performed on the HPC clusters “Clover” and “HIMster2” at the Helmholtz Institute Mainz, and “Mogon 2” at Johannes Gutenberg-Universität Mainz. We are grateful to our colleagues of the CLS initiative for sharing gauge field configurations.
%
\bibliographystyle{JHEP}
\bibliography{mybibfile}

\end{document}